\documentclass[11pt]{article}
\usepackage{amssymb,amsmath,fullpage,times,epsf}
\usepackage{mathrsfs}
\usepackage{amssymb}
\usepackage{graphics}
\usepackage{graphicx}
\usepackage{subfigure}
\usepackage{float}
\usepackage{bm}
\usepackage{amsfonts}
\usepackage{color}

\allowdisplaybreaks

\newcommand{\me}{\mathrm{e}}
\newcommand{\mi}{\mathrm{i}}

\def\mb{\mathbf}
\def\ra{\rightarrow}

\def\lim{\mathrm{lim}}

\def\arg{\mathrm{arg}}

\def\PT{$\mathcal{P}\mathcal{T}$}

\def\eref#1{(\ref{#1})}

\usepackage[square,numbers,sort&compress]{natbib}
\def\Im{\mathrm{Im}}

\begin{document}
\date{}

\title{Darboux transformation and analytic solutions of the discrete \PT-symmetric  nonlocal nonlinear Schr\"{o}dinger equation}

\author{Tao Xu$^{1,}$\thanks{Corresponding
author, e-mail: xutao@cup.edu.cn}\,, Hengji Li$^{1}$, Hongjun Zhang$^{1}$, Min Li$^{2}$, Sha Lan$^{1}$
 \\{\em 1. College of Science, China University of Petroleum, Beijing 102249, China}
\\{\em 2.  Department of Mathematics and Physics, }\\
{\em  North China Electric Power University, Beijing 102206, China}}\maketitle

\begin{abstract}
In this letter, for the
discrete parity-time-symmetric nonlocal nonlinear Schr\"{o}dinger equation, we construct the  Darboux transformation, which provides an algebraic iterative algorithm  to obtain a series of analytic solutions from a known one. To illustrate,  the breathing-soliton solutions,  periodic-wave solutions and localized rational soliton solutions are derived with the zero and plane-wave solutions as the seeds. The properties of those solutions are also discussed, and particularly the asymptotic analysis reveals all possible  cases of the interaction between the discrete
rational dark and antidark solitons.
\vspace*{4mm}


\end{abstract}

\section{Introduction}

In 1998, Bender and Boettcher first pointed out that a
non-Hermitian Hamiltonian can have the real and positive
eigenvalues  provided that it meets the combined parity and time
reversal symmetry (usually called the $\mathcal{P}\mathcal{T}$
symmetry)~\cite{bender1}. Since then, the non-Hermitian but \PT-symmetric systems have appeared  in many areas such as nonlinear optics~\cite{NO}, complex
crystal~\cite{crystal}, quantum chromodynamics~\cite{Markum},
Bose-Einstein condensates~\cite{BEC} in addition to
quantum mechanics~\cite{bender1}. Also, the notion of $\mathcal{P}\mathcal{T}$ symmetry has been extended to nonlinear integrable systems in mathematical physics~\cite{Ablowitz1,Sarma,Khare,LiXuPRE,XuLiJPSJ,Sarma2,Yan1,Fokas,Ablowitz3}. It should be noted that exactly solvable integrable systems are ubiquitous in nonlinear science
and play an important role in describing various nonlinear wave phenomena, like solitary waves~\cite{SW} and rogue waves~\cite{Roguewaves}.

In 2013,
Ablowitz and Musslimani  proposed the following continuous  \PT-symmetric   nonlocal nonlinear Schr\"{o}dinger (NNLS) equation~\cite{Ablowitz1}:
\begin{align}
\mi\,q_t(x,t)=q_{xx}(x,t)\,\pm\,2\,q(x,t)q^*(-x,t)q(x,t),
\label{CNLS}
\end{align}
where $*$ denotes complex conjugation,  the nonlinear term is nonlocally dependent on the
values of $q$ both at $x$ and $-x$,   and the  self-induced potential $V(x,t)=\pm\,2\,q(x,t)q^*(-x,t)$ obey the \PT-symmetric  condition $V(x,t)=V^*(-x,t)$.  Eq.~\eref{CNLS} is integrable
in the sense that it admits the Lax pair and an infinite number of conservation laws, and thus its initial-value problem can be solved by the inverse scattering transform (IST)~\cite{Ablowitz1}. Recently, much effort has been made to construct the exact analytic solutions of Eq.~\eref{CNLS}, including the singular pure soliton solution on the vanishing background~\cite{Ablowitz1},  static bright and dark soliton solutions~\cite{Sarma,Khare}, periodic and hyperbolic soliton solutions~\cite{Khare}, exponential and rational soliton solutions on the continuous wave background~\cite{LiXuPRE,XuLiJPSJ}, and   Peregrine-type rogue waves on the finite background~\cite{Sarma2}. In addition, the integrable vector extension and two-dimensional generalization of Eq.~\eref{CNLS} have also been discussed~\cite{Yan1,Fokas}.

In this letter, we will study the
discrete \PT-symmetric NNLS equation~\cite{Ablowitz3}:
\begin{align}
\mi\,\frac{dQ_n}{dt}=Q_{n+1}-2\,Q_n+Q_{n-1}-\epsilon\,Q_n Q_{-n}^*(Q_{n+1}+Q_{n-1}),
 \label{DNLS}
\end{align}
which is also an integrable Hamiltonian model~\cite{Ablowitz3}, where $Q_n(t)$ is a
complex-valued function, $n$ is an integer, $\epsilon=\mp 1$
denotes the focusing  and defocusing cases, respectively. The IST scheme of solving Eq.~\eref{DNLS} has been established~\cite{Ablowitz3}, and some singular soliton solutions and periodic-wave solutions of Eq. (2) with $ \epsilon=-1$ have also been obtained by the Hirota method~\cite{Zhu}. The main work in this study is to construct the $N$-time iterated Darboux transformation (DT), which provides an algebraic iterative algorithm to obtain the analytic solutions of Eq.~\eref{DNLS} from a known one. To illustrate, with the zero and plane-wave solutions as the seeds, we derive the breathing-soliton solutions,  periodic-wave solutions and localized rational soliton solutions. In particular, via asymptotic analysis we reveal three types of elastic interactions between  rational dark (RD) and rational antidark (RAD)  solitons, and two types of degenerate two-soliton interactions.

\section{Darboux transformation of Eq.~\eref{DNLS}}

The Lax pair of Eq.~\eref{DNLS} takes the form
\begin{subequations}
\begin{align}
&\Phi_{n+1}=U_n\Phi_n,  \quad
U_n=\begin{pmatrix}
\ z & Q_n \\
R_n & z^{-1}
\end{pmatrix} , \label{DNLS2a} \\
&\frac{d\Phi_n}{dt}=V_n\Phi_n, \quad
V_n=\begin{pmatrix}
\mi\, R_{n-1}\, Q_n+\mi(1-z^2) & -\mi \,z\,Q_n+\mi\,z^{-1} Q_{n-1}\\
-\mi \,z R_n+\mi\, z^{-1}R_{n-1} & -\mi\, Q_{n-1}\, R_n-\mi(1-z^{-2})
\end{pmatrix},\label{DNLS2b}
\end{align}\label{DNLS2}
\end{subequations}
with $R_n=\epsilon\,Q_{-n}^*$, where $ \Phi_n=(\phi_{1,n}, \phi_{2,n})^{\rm{T}}$
is the vector eigenfunction, $z$ is a complex spectral parameter.
 As a special gauge transformation, the DT comprises of the eigenfunction and potential transformations. For the once-iterated DT, we take the  eigenfunction transformation on System~\eref{DNLS2} be of the form
\begin{align}
\Phi_{n}^{[1]}=T_{n}^{[1]}\Phi_{n}, \quad   T_{n}^{[1]}=
\begin{pmatrix}
 m_1(t) \left(z- a_n(t) z^{-1}\right) & -b_n(t)  \\
  -c_n(t) & h_1(t) \left(z^{-1}- d_n(t)z\right)   \\
\end{pmatrix}, \label{eigenfunctionTran}
\end{align}
where $T_{n}^{[1]}$ is the once Darboux matrix, $m_1(t)$, $h_1(t)$,  $a_n(t)$,
$b_n(t)$, $c_n(t)$ and $d_n(t)$  are to be
determined, and $\Phi_{n}^{[1]}=(\phi^{[1]}_{1,n}, \phi^{[1]}_{2,n})^{\rm{T}}$ is the
once iterated eigenfunction.

From the knowledge of DT, $\Phi^{[1]}_{n}$ is required to
satisfy 
\begin{align}
& T_{n+1}^{[1]} U_{n}= U_{n}^{[1]}T_{n}^{[1]}, \quad
\frac{d}{d t} T_n^{[1]} + T_{n}^{[1]}V_n = V_{n}^{[1]}T_{n}^{[1]}, \label{DTinvariance}
\end{align}
where $U^{[1]}_{n}$ and $V^{[1]}_{n}$ are the same as $U_n$ and $V_n$
except that $Q_n$ and $Q^*_{-n}$ are replaced by
$Q_{n}^{[1]}$ and $Q_{-n}^{*[1]}$,  respectively.
We note the fact~\cite{Ablowitz3} that if $\Phi_{1,n}=\big(f_{1,n},
g_{1,n})^{\rm{T}}$ satisfies System~\eref{DNLS2} with $z=z_1$, then
$\bar{\Phi}_{1,n}=w_{-n}^*\big(g^*_{1,1-n}, $ $-\epsilon f^*_{1,1-n}\big)^{\rm{T}}$ is also a solution of System~\eref{DNLS2} with $z = z^*_1$, where $w_n=\prod\limits_{k=-\infty}^{n}\frac{1}{1-\epsilon\, Q_k Q_{-k}^*}$.
Thus,  $a_n$, $b_n$, $c_n$ and $d_n$ can be determined by 
demanding $ T_{n}^{[1]}\mid_{z=z_1}\Phi_{1,n}=\mb{0}$ and $ T_{n}^{[1]}\mid_{z=z^*_1} \bar\Phi_{1,n}=\mb{0}$.
Furthermore, one can check that
Eq.~\eref{DTinvariance} is satisfied if
$Q_{n}^{[1]}$ and $Q_{-n}^{*[1]}$ are given by
\begin{align}
Q_n^{[1]}=-\frac{m_1 Q_n+b_n}{h_1 d_n }, \quad
R_{n}^{[1]}=-\frac{c_n+  h_1 R_n}{m_1 a_n}, \label{Potentialtransform}
\end{align}
where the symmetry reduction $R^{[1]}_n = \epsilon\,Q^{*[1]}_{-n} $ holds if taking
$h_1(t)=\vert z_1\vert$ and $m_1(t)=\vert z_1\vert^{-1} $.
Next, we construct the $N$-time iterated DT of Eq.~\eref{DNLS}. In doing so, we successively implement  the once-iterated eigenfunction transformation, that is,
\begin{align}
&\Phi_{n}^{[N]}= T_{N,n}^{[1]}\cdots T_{1,n}^{[1]} \Phi_{n},  \quad
 T^{[1]}_{k,n}=
\begin{pmatrix}
 m_k(t) \left(z- a_{k,n}(t) z^{-1}\right) & -b_{k,n}(t)  \\
  -c_{k,n}(t) & h_k(t) \left(z^{-1}- d_{k,n}(t) z\right)   \\
\end{pmatrix}, \label{NDT}
\end{align}
where  $1\leq k\leq N$, $\Phi_{n}^{[N]}=(f_{1,n}^{[N]},g_{1,n}^{[N]})^{\rm{T}}$ is the
$N$-time iterated eigenfunction, $h_k(t)=\vert z_k\vert$ and $m_k(t)=\vert z_k\vert^{-1}$.
The $N$-time iterated Darboux matrix $T_{n}^{[N]}$ can be written as
\begin{align}
T_{n}^{[N]} = T_{N,n}^{[1]}\cdots T_{1,n}^{[1]} =
\begin{pmatrix}
A_N(n,t,z)  & B_N(n,t,z)  \\
C_N(n,t,z)  & D_N(n,t,z)   \\
\end{pmatrix},  \label{eigenfunctionTran1}
\end{align}
with $ A_N = \mathfrak{M}\,z^N  + \sum\limits_{j=1}^{N}a_{n}^{(j)}(t) z^{N-2j}$, $B_N =  \sum\limits_{j=1}^{N}b_n^{(j)}(t) z^{N+1-2j}$, $C_N=   \sum\limits_{j=1}^{N}c_n^{(j)}(t) z^{N+1-2j}$,
$D_N= \mathfrak{M}^{-1}z^{-N} + \sum\limits_{j=1}^{N}d_n^{(j)}(t) z^{N+2-2j}$ and $\mathfrak{M} =\prod\limits_{k=1}^{N}|z_{k}|^{-1}$.

Similarly,  $a_n^{(j)}(t)$,
$b_n^{(j)}(t)$, $c_n^{(j)}(t)$ and $d_n^{(j)}(t)$ ($1\leq j\leq N$) can be
uniquely determined by requiring $T_{n}^{[N]}\mid_{z=z_k}\Phi_{k,n}=\mb{0}$ and $
T_{n}^{[N]}\mid_{z=z^*_k} \bar\Phi_{k,n}=\mb{0} $  $(1\leq k\leq N)$,
where $\Phi_{k,n}=\big(f_{k,n},
g_{k,n})^{\rm{T}}$ and $\bar{\Phi}_{k,n}=w_{-n}^*\big(g^*_{k,1-n}, -\epsilon f^*_{k,1-n} \big)^{\rm{T}}$ ($w_{n}=\prod\limits_{k=-\infty}^{n}\frac{1}{1-\epsilon Q_k Q_{-k}^*}$) are,
respectively, the solutions of System~\eref{DNLS2} with $z=z_k$ and  $z = z^*_k$. Via Cramer's rule,
the $N$-time iterated potentials $Q_n^{[N]}$ and $R_{n}^{[N]}$  can be represented as
\begin{align}
 Q_n^{[N]}= \frac{Q_n \mathfrak{M}-b_n^{(1)}}{d_n^{(1)}},  \quad  R_{n}^{[N]}= \frac{R_n \mathfrak{M}^{-1}-c_n^{(N)}}{a_n^{(N)}}, \label{NPotentialTranb}
\end{align}
with
\begin{align}
& a_n^{(N)}=(-1)^{N}\mathfrak{M}\frac{\tau_n(N,N-2;N-1,N-1)}{\tau_n(N-2,N;N-1,N-1)},  \,\, b_n^{(1)}=(-1)^{N+1}\mathfrak{M}\frac{\tau_n(N,N;N-3,N-1)}{\tau_n(N-2,N;N-1,N-1)},   \notag \\
& c_n^{(N)}=(-1)^{N+1}\mathfrak{M}^{-1}\frac{\tau_n(N-1,N-3;N,N)}{\tau_n(N-1,N-1;N,N-2)},  \,\,
 d_n^{(1)}=(-1)^{N}\mathfrak{M}^{-1}\frac{\tau_n(N-1,N-1;N-2,N)}{\tau_n(N-1,N-1;N,N-2)}, \notag
\end{align}
where the determinant $\tau_n(M,L;M',L')$ is defined as
\begin{align}
& \tau_n(M,L;M',L')=\begin{vmatrix}
F_{n}(M,L)&  G_{n}(M',L')\\
G_{1-n}^*(M,L)&  -\epsilon F_{1-n}^*(M',L')\\
\end{vmatrix},\quad d_n^{(1)}=a_n^{(N)}\mathfrak{M}^{4},
\end{align}
with $ F_{n}(M,L) = \big(z_k^{M+2-2j}f_{k,n}\big)_{\substack{1 \leqslant k \leqslant  N, \\ 1 \leqslant j
\leqslant  \frac{M+L}{2}+1 }}$,  $G_{n}(M',L') = \big(z_k^{M'+2-2j}g_{k,n}\big)_{\substack{1 \leqslant k \leqslant  N, \\ 1 \leqslant j\leqslant  \frac{M'+L'}{2}+1 }}$, $F_{1-n}^*(M',L') = \big(z_k^{*M'+2-2j}f_{k,1-n}^*\big)_{\substack{1 \leqslant k \leqslant  N, \\ 1 \leqslant j \leqslant  \frac{M'+L'}{2}+1 }}$, $G_{1-n}^*(M,L) = \big(z_k^{*M+2-2j}g_{k,1-n}^*\big)_{\substack{1 \leqslant k \leqslant  N, \\ 1 \leqslant j \leqslant  \frac{M+L}{2}+1 }}$.
Finally, it can be  proved that the  $N$-time iterated potentials in Eq.~\eref{NPotentialTranb} obey the reduction relation
$R^{[N]}_n = \epsilon Q^{*[N]}_{-n} $ (see Appendix A). 
Hence, we can safely say that Transformations~\eref{NDT} and~\eref{NPotentialTranb} constitute the $N$-time  iterated DT of Eq.~\eref{DNLS}.

\section{Analytic solutions on the vanishing and plane-wave backgrounds}

In this section, based on the above-obtained  DT algorithm,   we will choose the zero solution and plane-wave solutions as the seeds to derive some new analytic solutions of Eq.~\eref{DNLS}, including the breathing-soliton solutions,  periodic-wave solutions and localized rational soliton solutions.

\vspace*{2mm}
\noindent \textbf{A. Breathing-soliton solutions on the vanishing background}

\begin{figure}[H]
\begin{minipage}[t]{0.33\linewidth} 
\centering
\includegraphics[width=1.9in]{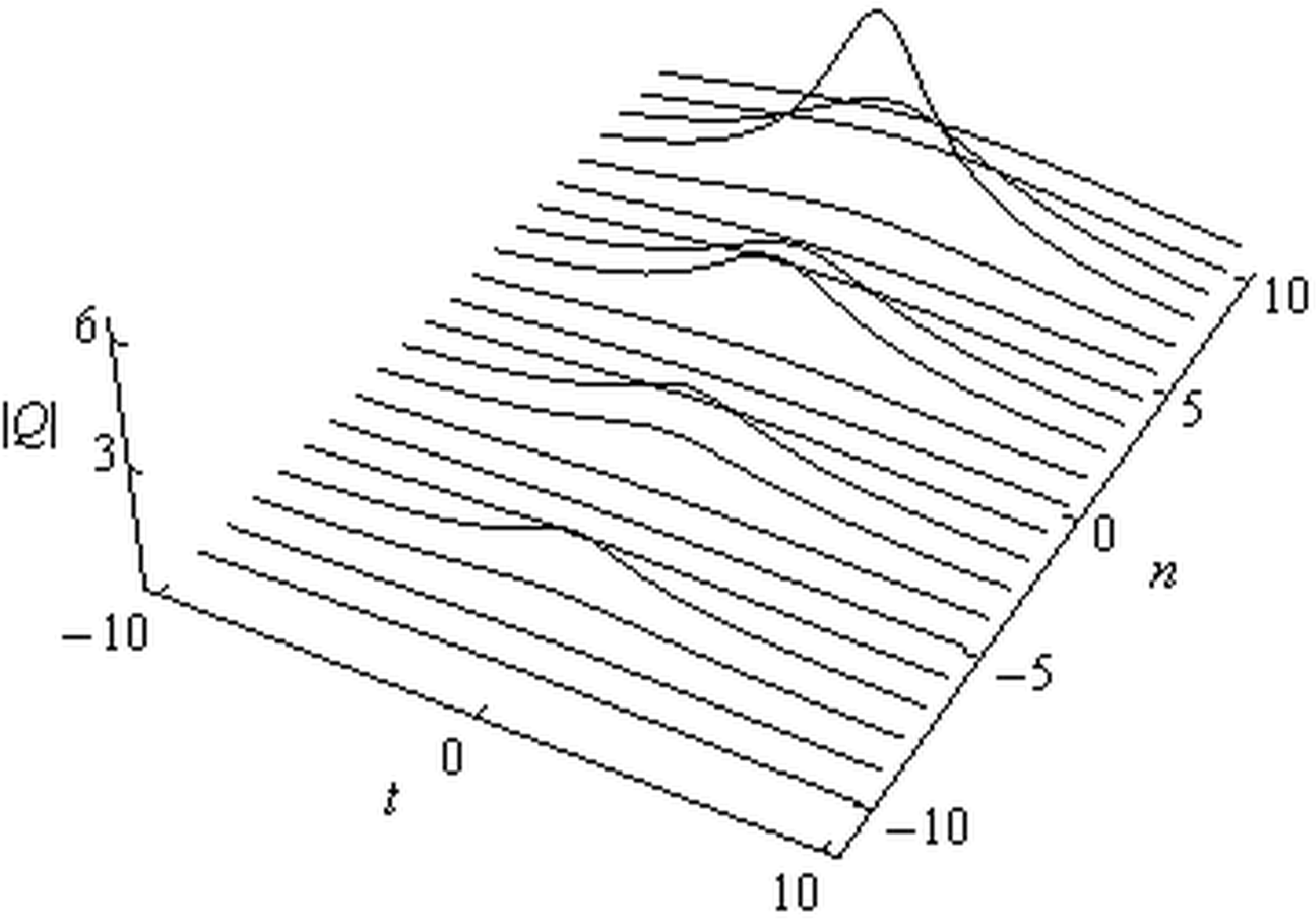}
\vspace*{6mm}
 \caption{\small Breathing one-soliton solution with $\epsilon=1$ and $z_1=1+\frac{\mi}{3}$.  \label{breather}}
\end{minipage}\hspace{5mm}
\begin{minipage}[t]{0.66\linewidth}
\centering
\subfigure[]{\label{breather2}
\includegraphics[width=1.9in]{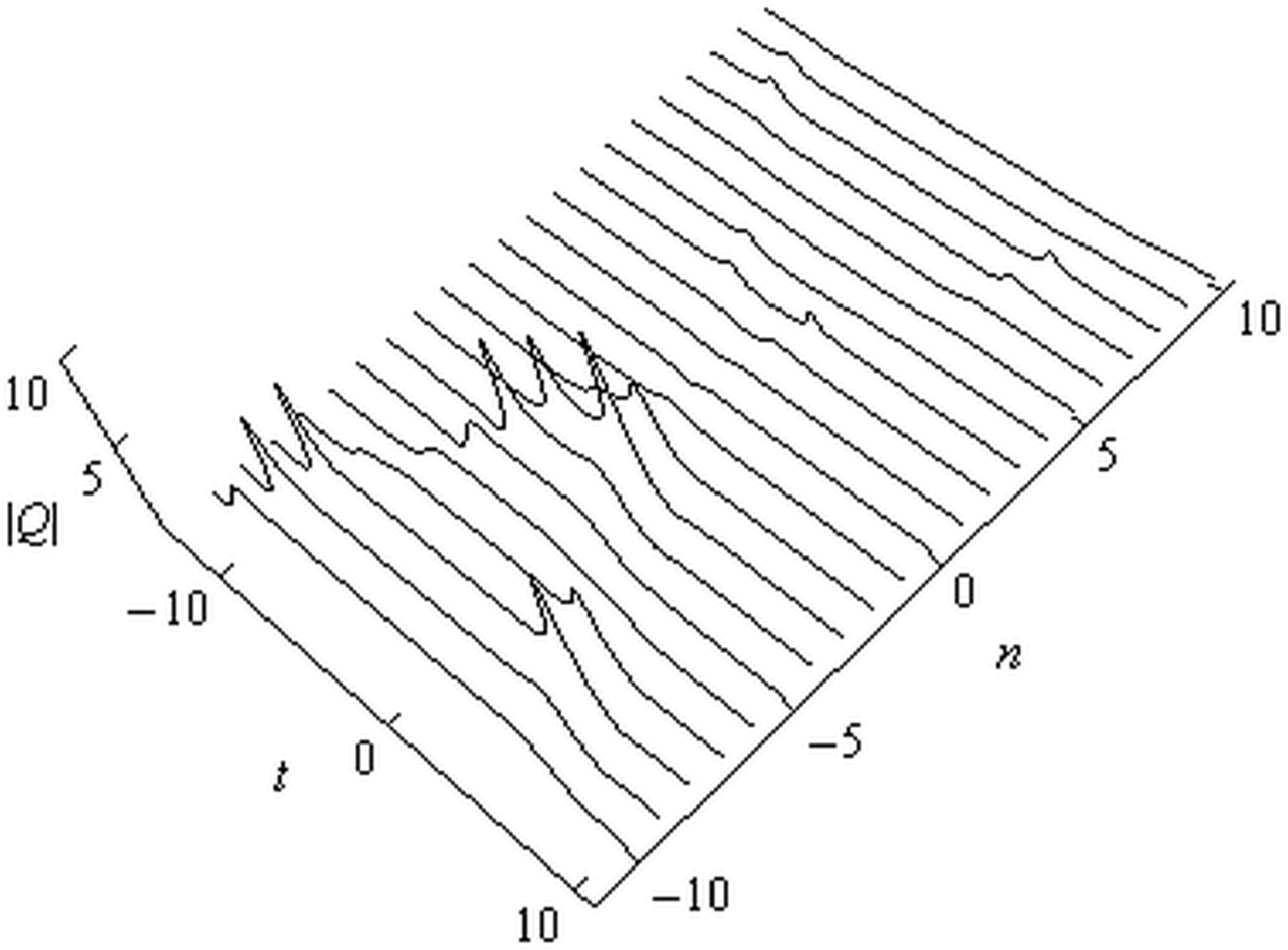}}\hfill
\subfigure[]{ \label{periodic}
\includegraphics[width=1.9in]{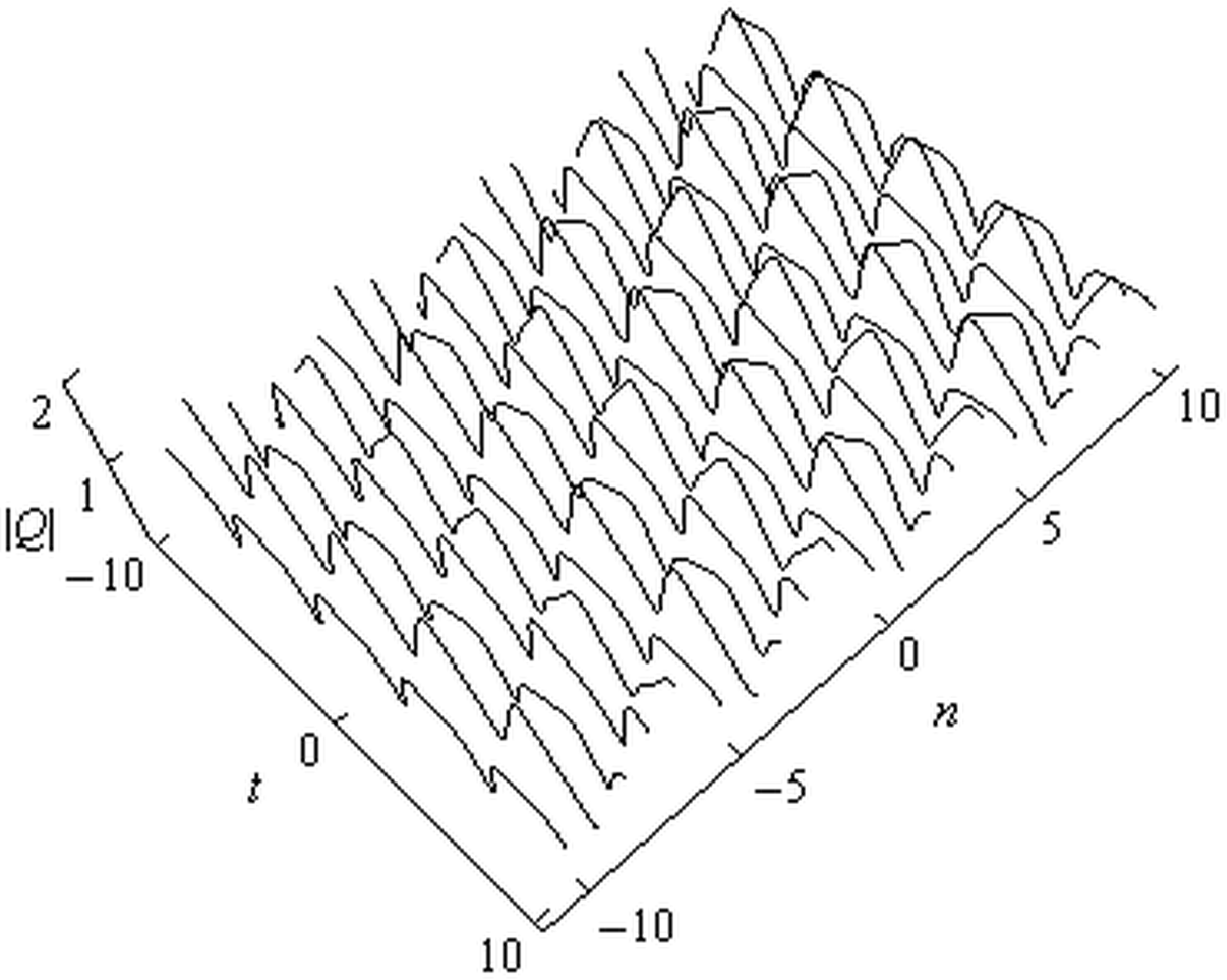}}
\caption{\small  (a) Breathing-soliton   and (b) periodic-wave solutions on the plane-wave background, where
$c= \frac{1}{2}$, $\epsilon = 1$, $\alpha_1= 1$, (a) $\beta_1= 1+2\,\mi$, $z_1= -\frac{3}{5}-\frac{\mi}{5}$, and (b) $\beta_1= 1$, $z_1= \frac{\sqrt{2}}{2}+\frac{\sqrt{2}\,\mi}{2}$.  }
\end{minipage}
\end{figure}

First, starting from the seed  $Q_n=0$, we solve System~\eref{DNLS2} with $Q_n=0$ and $z=z_1$, yielding
\begin{align}(f_{1,n}, g_{1,n}) = \big(\alpha_1 z_1^n \me^{\mi t(1-z_1^2)}, \beta_1 z_1^{-n} \me^{-\mi t(1-z_1^{-2})}\big),
\label{LPsolutions}
\end{align}
where $ \alpha_1$ and $\beta_1 $ are two
nonzero complex constants.  Substituting~\eref{LPsolutions} into Eq.~\eref{Potentialtransform} gives rise to
\begin{align}
Q_n^{[1]}=\frac{\gamma^*_1  (z^*_1)^{2n-1} (z_1^{*2} - z_1^2)\me^{-\mi \chi_1^{*2}\,t }}{\epsilon |z_1|^2 +|\gamma_1|^2 \big(\frac{z_1^*}{z_1}\big)^{2n}\me^{\mi\,(\chi_1^2- \chi_1^{*2})\,t} }, \quad (\chi_1= z_1-z_1^{-1},\,\, \gamma_1=\frac{\beta_1}{\alpha_1}),  \label{per1}
\end{align}
which has also been obtained by the
IST method~\cite{Ablowitz3} and by the Hirota method~\cite{Zhu}.
Note that this solution is singular in the continuous limit. However, because $n$ is an integer variable,  all the singularities can be avoided if taking  $\arg(z_1) \neq \frac{m\,\pi}{4\,n}\,\,(m,n\in\mathbb{Z})$ for  $|z_1| = 1$ and $|\gamma_1| \neq 1$ for $|z_1| \neq 1$.  In illustration, Fig.~\ref{breather} presents a discrete breathing soliton which  oscillates periodically along the $n$ axis.


\noindent \textbf{B. Breathing-soliton and periodic-wave solutions on the plane-wave background}

Second, we try to construct the soliton
solutions on the nonvanishing background. It is easy to find that
Eq.~\eref{DNLS} admits the plane-wave solution $ Q_n=c\,\me^{2\,\mi \epsilon |c|^2 t }$, where $c$ is a complex constant. Then, we substitute
the solution into System~\eref{DNLS2}  and solve the resulting equations with $z=z_1$.
If $c$ and $z_1$ satisfy the condition $ \chi^2_1 + 4\,\epsilon |c|^2 \neq 0$ ($\chi_1= z_1-z_1^{-1}$),
one can obtain
\begin{align}\label{laxpairsolution}
\begin{pmatrix}
f_{1,n}\\
g_{1,n}
\end{pmatrix}=
\begin{pmatrix}
\me^{\mi \epsilon |c|^2 t }(\alpha_1 \me^{ n \ln{\mu_1^{-}}-\mi \chi_1\mu_1^{-}t}
+\beta_1 \me^{n \ln{\mu_1^{+}}-\mi \chi_1 \mu_1^{+}t})\\
\frac{\me^{-\mi \epsilon |c|^2 t }}{c}\big[(z_1-\mu_1^{-}) \alpha_1 \me^{n \ln{\mu_1^{-}}-\mi \chi_1 \mu_1^{-} t}
+(\mu_1^{+}-z_1) \beta_1 \me^{n \ln{\mu_1^{+}}-\mi \chi_1 \mu_1^{+} t}\big]
\end{pmatrix},
\end{align}
where $ \mu_1^{\pm}=\frac{z_1^2+1\pm\sqrt{(z^2_1-1)^2+4\epsilon |c|^2z_1^2}}{2z_1}$,
and $ \alpha_1$ and $\beta_1 $  are two
nonzero complex parameters. Then, with substitution of~\eref{laxpairsolution} into Eq.~\eref{Potentialtransform}, we have the breathing-soliton solution (which is omitted here because  its expression is too long and complicated) on the plane-wave background. With suitable choice of the involved parameters, the solution has no singularity for all integers $n$. But different from the exponential dark and antidark  solitons in the defocusing case of Eq.~\eref{CNLS}~\cite{LiXuPRE}, there appear two breathing solitons under the nonsingular condition and they exhibit the elastic interactions on the background [see Fig.~\ref{breather2}].
In particular, with $|z_1|=1$ the solution reduces to the triangular periodic solution and  describes the interaction of periodic waves [see Fig.~\ref{periodic}].

\begin{figure}[H]
 \centering
\subfigure[]{\label{Fig1a}
\includegraphics[width=2in]{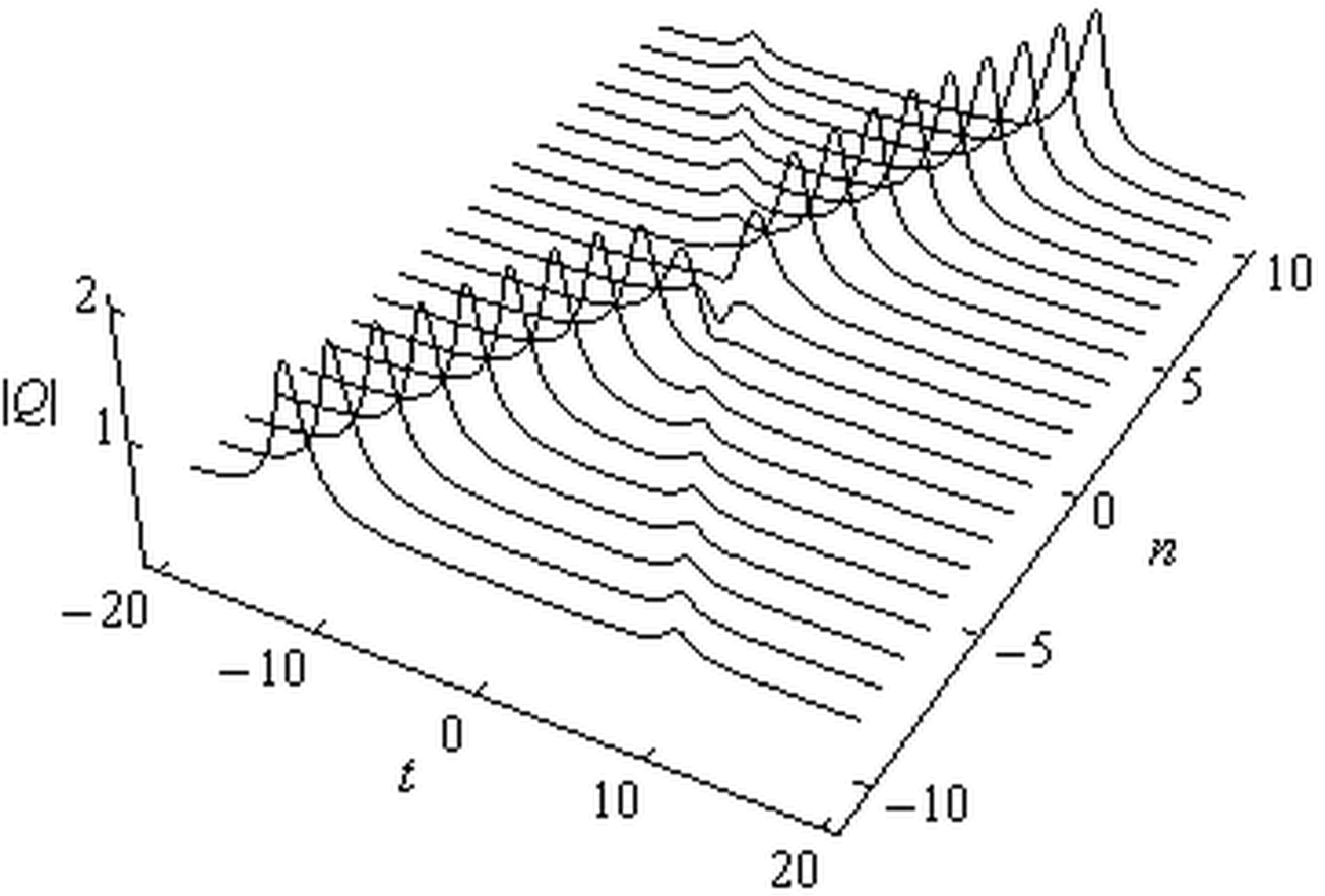}}\hfill
\subfigure[]{ \label{Fig1b}
\includegraphics[width=2in]{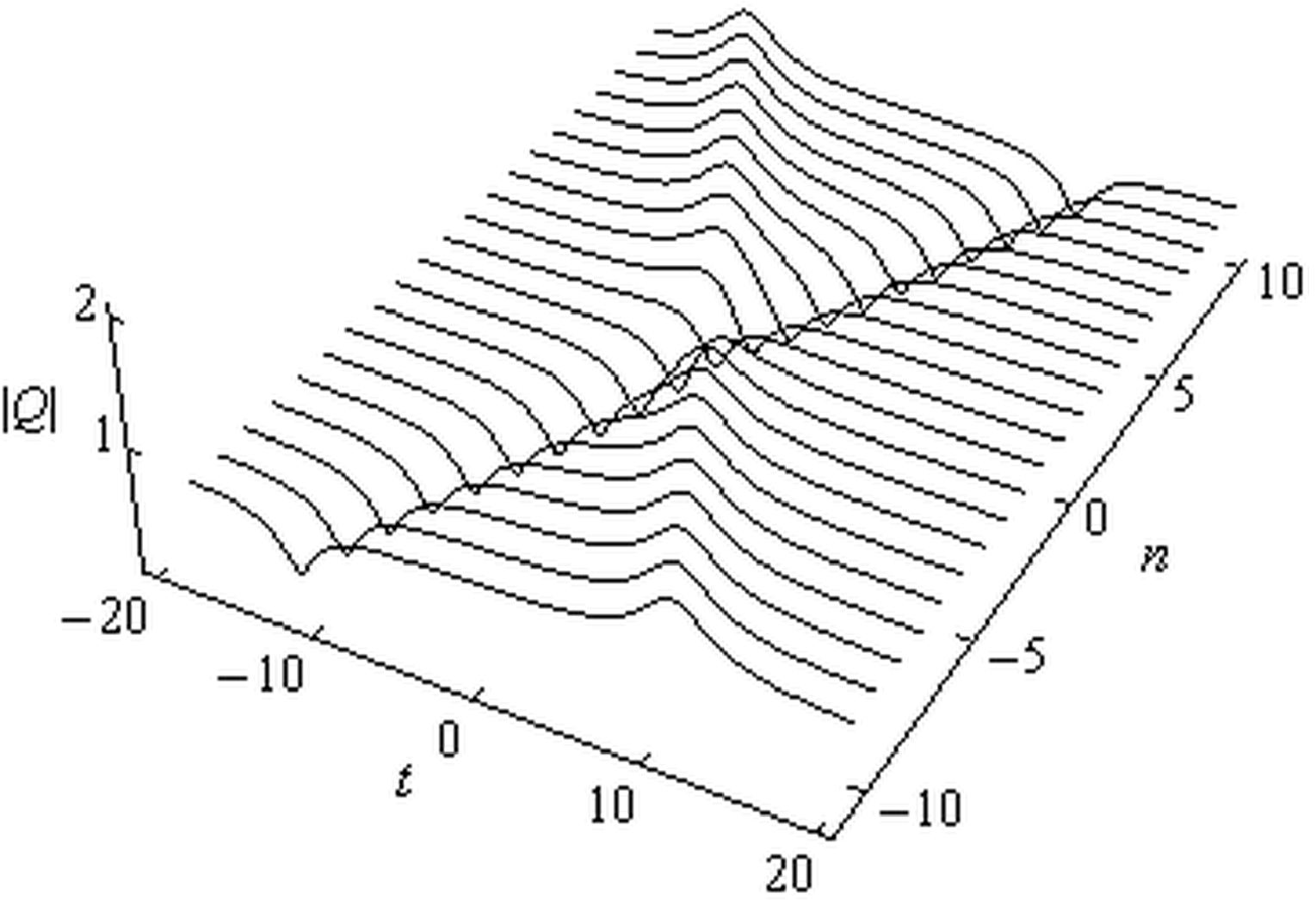}}\hfill
\subfigure[]{ \label{Fig1c}
 \includegraphics[width=2in]{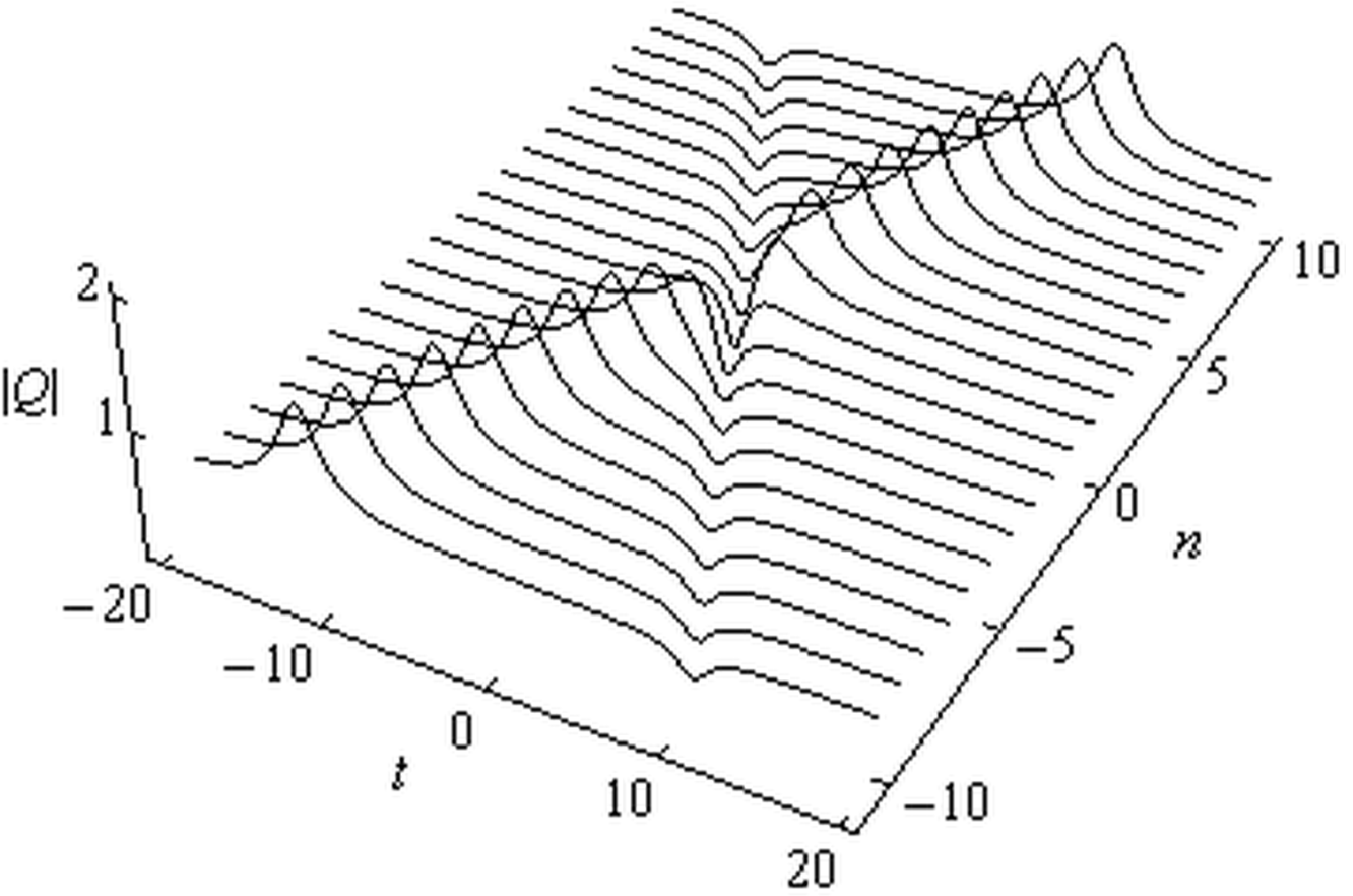}}
\caption{\small  Three types of elastic two-soliton interactions with $c=\frac{1}{2}$ and $\theta=\frac{\pi}{6}$, where
(a) RAD-RAD interaction ($\gamma_1=1+1.6\,\mi$), (b) RAD-RD interaction ($\gamma_1=1-\,\mi$), and (c) RD-RAD interaction ($\gamma_1=1+2\,\mi$).}
\end{figure}
\noindent \textbf{C. Localized rational soliton solutions on the plane-wave background}

Third, we also choose the seed solution $Q_n=c\,\me^{2\,\mi \epsilon |c|^2 t }$ but consider the particular case
$ \chi^2_1 + 4\,\epsilon |c|^2  = 0$. Note that if $\epsilon=-1$ or $ |c| \ge 1$,  $z_1$ reduces to a pure real or imaginary number, so that the DT is trivial and  no new solution can be generated.  Thus, we must impose
$ \epsilon=1$ and $|c|<1$, and  obtain that $ z_1=\me^{\mi \theta}$ ($\tan\theta=\frac{\pm |c|}{\sqrt{1-|c|^2}}$).
In this case, System~\eref{DNLS2}  has the following solution:
\begin{align}
\begin{pmatrix}
 f_{1,n} \\
 g_{1,n} \\
\end{pmatrix}
=\begin{pmatrix}
 \frac{\mi  \sin\theta}{c^*}\me^{\mi|c|^2t} \cos^n\!\theta\cdot\me^{t  \sin (2\theta)}
[n+t \sin (2\theta)+\gamma_1-\mi \cot\theta])\\
\me^{-\mi|c|^2t} \cos^n\!\theta\cdot\me^{t \sin (2\theta)} [n+t \sin (2\theta)  +\gamma_1]\\
\end{pmatrix}, \label{LPSol3}
\end{align}
where $\gamma_1$ is a  nonzero complex constant.
Substituting Eq.~\eref{LPSol3} into Eq.~\eref{Potentialtransform}, we have
\begin{align}
Q_n^{[1]}=c\,\me^{2 \mi |c|^2 t}  \frac{(\xi-K-\mi \cot\theta)(\eta+K^*+\mi \cot\theta)-\frac{\csc^2\theta}{4 }}
{ \xi\, \eta -K\eta +K^*\xi-|K|^2-\frac{\csc^2\theta}{4 }}, \label{solution}
\end{align}
with $\xi=n+\sin(2 \theta)\cdot t $, $\eta=n-\sin(2 \theta)\cdot t $, $K=\frac{\mi \cot\theta
-1}{2}-\gamma _1$.  It can be proved that Solution~\eref{solution} has no singularity if and only if $
\Im(\gamma_1)\neq \frac{\cot\theta}{2}$.
\begin{figure}[H]
\centering
\subfigure[]{\label{Fig2a}
\includegraphics[width=2in]{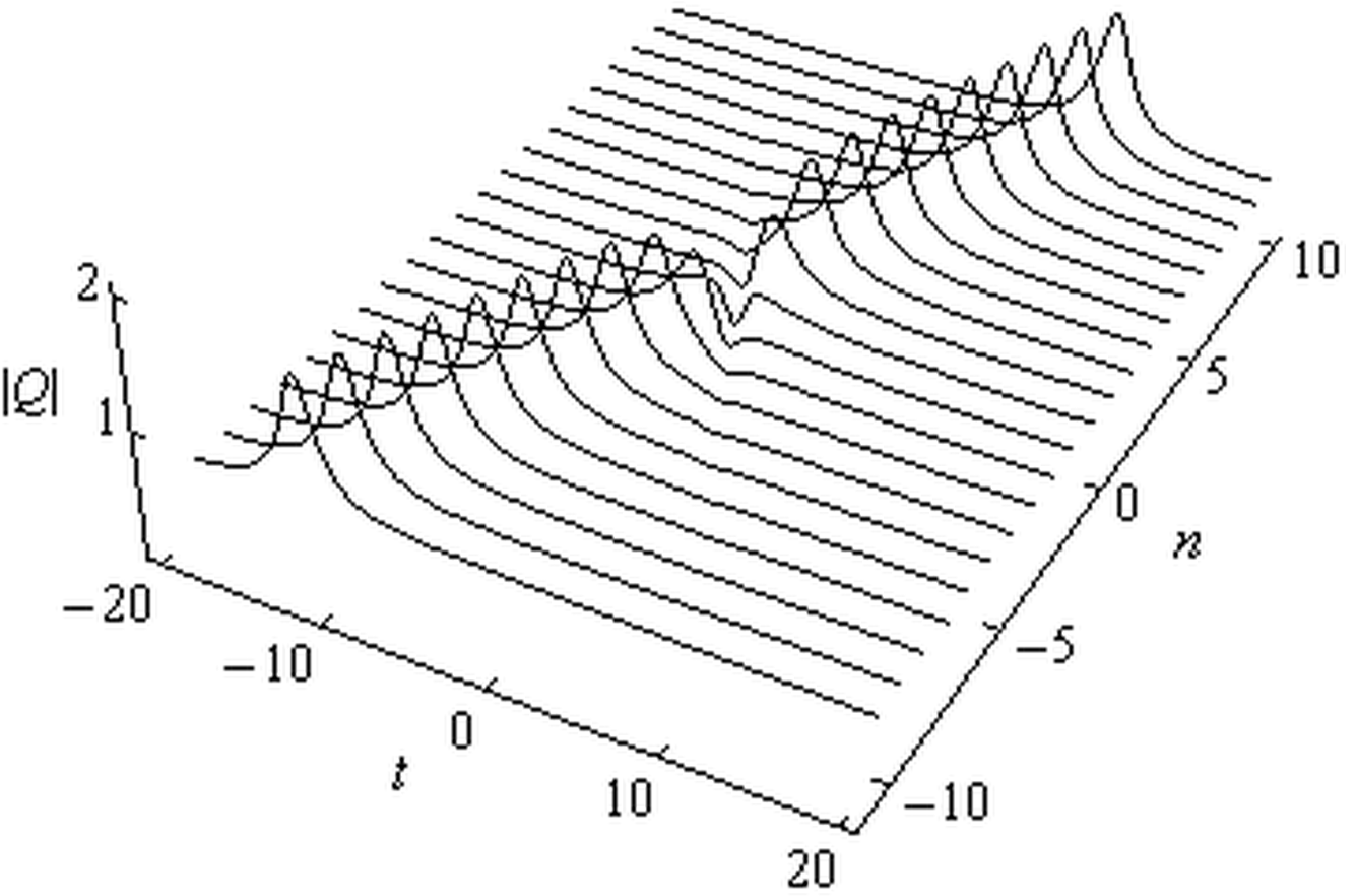}}\hspace{2.5cm}
\subfigure[]{ \label{Fig2b}
\includegraphics[width=2in]{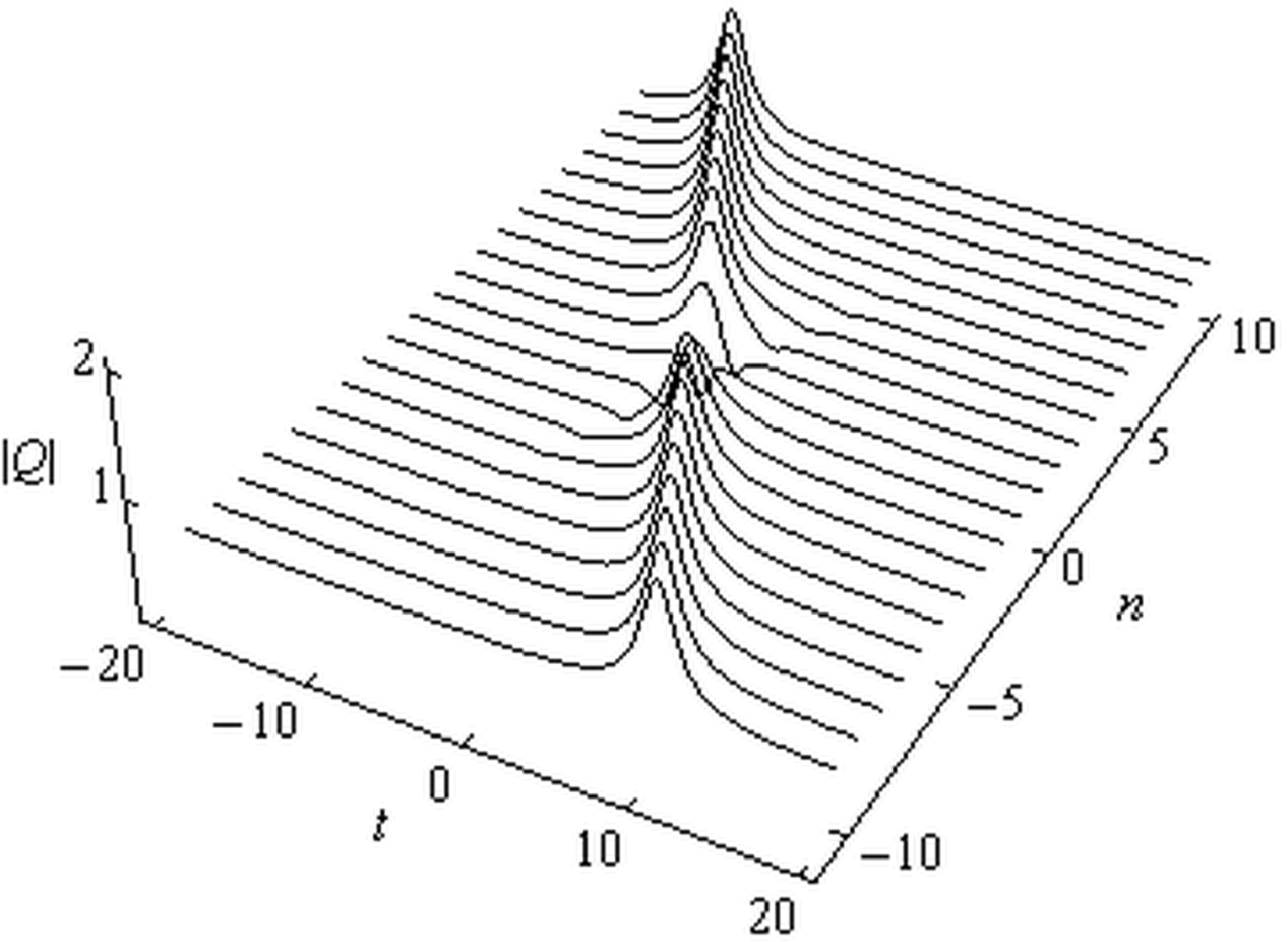}}
\caption{\small  Degenerate two-soliton interactions with $c=\frac{1}{2}$, $\theta=\frac{\pi}{6}$,
(a) $\gamma_1=1+\sqrt{3}\,\mi$ and (b) $\gamma_1=1$.}
\end{figure}

When $|t| \rightarrow \infty$, we obtain two asymptotic expressions of Solution~\eref{solution} as follows:
\begin{align}
 Q_n^{[1]}\rightarrow\left\{
\begin{array}l
 S_n^{(\rm{I})}=c \,\me^{2 \mi |c|^2 t } (1+\frac{\mi \cot \theta}{K-\xi}), \quad n+\sin(2 \theta)\cdot t  \sim 0, \\
 S_n^{(\rm{I\!I})}=c\,  \me^{2 \mi |c|^2 t } (1+\frac{\mi \cot \theta}{K^*+\eta}),  \quad n-\sin(2 \theta)\cdot t  \sim 0,
\end{array}
\right. \label{ij1}
\end{align}
where $S_n^{(\rm{I})}$  describes an RD or RAD soliton for $1-\Im(\gamma_1)\tan\theta<0$ or $1-\Im(\gamma_1)\tan\theta>0$, while $S_n^{(\rm{I\!I})}$ represents
an RD or RAD soliton for $\Im(\gamma_1) \cot\theta<0$ or $\Im(\gamma_1) \cot\theta>0$. Similar to the
defocusing case occurring in Eq.~\eref{CNLS}~\cite{XuLiJPSJ}, one can obtain the elastic interactions of the discrete  RAD-RAD, RAD-AD and AD-RAD  soliton pair on a plane-wave background, and there is no phase shift for the interacting solitons [see Figs.~\ref{Fig1a}--\ref{Fig1c}]. Particularly when $1-\Im(\gamma_1)\tan\theta=0$ or $\Im(\gamma_1) \cot\theta=0$, the asymptotic soliton $S_n^{(\rm{I})}$ or $S_n^{(\rm{I\!I})}$ disappears for large values of $t$, as shown in Figs.~\ref{Fig2a} and~\ref{Fig2b}. The relevant parametric conditions for five different asymptotic patterns of Solution~\eref{solution} are presented in Table~\ref{Table}.
\vspace*{-5mm}
\begin{table}[h]
\caption{\small Asymptotic patterns of Solution~\eref{solution} under
different parametric conditions. \label{Table}} \vspace{0mm}
\begin{center}\small{
\begin{tabular}{|c|c|c|c|c|}
\hline  \multicolumn{1}{|c|}{Parametric conditions} & {Asymptotic
soliton $S_n^{(\rm{I})}$} & {Asymptotic soliton $S_n^{(\rm{I\!I})}$}
\\  \hline  $ 1-\Im(\gamma_1)\tan\theta>0 $, $ \Im(\gamma_1) \cot\theta
>0$   &  RAD soliton  &   RAD soliton
\\  \hline  $ 1-\Im(\gamma_1)\tan\theta>0 $, $ \Im(\gamma_1) \cot\theta
<0$   &  RAD soliton  &   RD soliton
\\  \hline  $ 1-\Im(\gamma_1)\tan\theta<0 $, $ \Im(\gamma_1) \cot\theta
>0$   &  RD soliton  &   RAD soliton
\\  \hline  $ 1-\Im(\gamma_1)\tan\theta=0 $    &  Vanish  &   RAD soliton
\\  \hline  $ \Im(\gamma_1) \cot\theta =0 $  &  RAD soliton  &   Vanish
\\ \hline
\end{tabular}}
\end{center}
\end{table}

\vspace{-5mm}

\section{Concluding remarks}

Currently, it has been an important concern to study the $\mathcal{P}\mathcal{T}$-symmetric  integrable systems in nonlinear mathematical physics. In this letter, for the
discrete \PT-symmetric NNLS equation~\eref{DNLS}, we have constructed its $N$-time iterated DT and have
represented the iterated solutions in terms of some simple determinant.
To illustrate, with the zero and plane-wave solutions as the seeds, we have derived the breathing-soliton solutions,  periodic-wave solutions and localized rational soliton solutions. Also, we have discussed the properties of those solutions, and particularly revealed the elastic interactions of the discrete RAD-RAD, RAD-AD and AD-RAD soliton pair on the plane-wave background. It should be mentioned that the defocusing Ablowitz-Ladik model admits the exploding rogue-wave solutions which develop singularity at a certain specific time~\cite{DAL}. Very differently, the rational solution~\eref{solution} has no singularity only if $\Im(\gamma_1)\neq \frac{\cot\theta}{2}$, and can exhibit the elastic interactions between two traveling rational solitons. In the future, it is worth to further study the stability of localized rational soliton solutions and the dynamical properties of the multi-iterated solutions via the DT.

\section*{Acknowledgments}
T. Xu would like to thank the financial support by  the Natural Science Foundation of Beijing,
China (Grant No.~1162003),   the Science Foundations of China
University of Petroleum, Beijing (Grant Nos. 2462015YQ0604 and
2462015QZDX02) and the National Natural Science Foundations of China
(Grant No.~11371371). M. Li  thanks the financial support by the  National Natural Science Foundations of China
(Grant Nos.~61505054 and 11426105).

\section*{Appendix A: Proof of the reduction relation $R^{[N]}_n = \epsilon Q^{*[N]}_{-n} $ in Eq.~\eref{NPotentialTranb}}
 \label{appendixA}
\renewcommand{\theequation}{A.\arabic{equation}}
\setcounter{equation}{0}

Recall that  $\Phi_{k,n}=\big(f_{k,n},
g_{k,n})^{\rm{T}}$ and $\bar{\Phi}_{k,n}=w_{-n}^*\big(g^*_{k,1-n}, -\epsilon f^*_{k,1-n} \big)^{\rm{T}}$ satisfy
System~\eref{DNLS2} with $z=z_k$ and  $z = z^*_k$, respectively. Thus,  we can arrive at the following identity relations:
\begin{align}
& \tau_n(N-1,N-1;N,N-2)=\mathfrak{M}^{-2}\tau_n(N-2,N;N-1,N-1), \label{A1}\\
& F_{n}(N-k,N-l)+Q_n G_{n}(N-k-1,N-l+1)=F_{n+1}(N-k,N-l),  \label{A2} \\
& \epsilon Q_{-n}^*F_{n}(N-k,N-l)+G_{n}(N-k-1,N-l+1)=G_{n+1}(N-k,N-l), \label{A3} \\
& G_{1-n}^{*}(N-k,N-l)-\epsilon Q_nF_{1-n}^{*}(N-k-1,N-l+1)\notag \\
&\hspace{1.5cm} =(1-\epsilon Q_n Q_{-n}^*)G_{-n}^{*}(N-k-1,N-l+1),  \label{A4}\\
& Q_{-n}^*G_{1-n}^{*}(N-k,N-l)-F_{1-n}^{*}(N-k-1,N-l+1),\notag\\
&\hspace{1.5cm} =(\epsilon Q_n Q_{-n}^*-1)F_{-n}^{*}(N-k,N-l).  \label{A5}
\end{align}
By virtue of the above identities, $Q^{[N]}_n$ and  $R^{[N]}_n$ in Eq.~\eref{NPotentialTranb} can be simplified as
{\small{\begin{align}
Q_n^{[N]}= &\frac{Q_n \tau_n(N-2,N;N-1,N-1)+(-1)^N\tau_n(N,N;N-3,N-1)}{(-1)^N\tau_n(N-1,N-1;N-2,N)}\notag\\
=&\frac{\begin{vmatrix}F_{n+1}(N-1,N-1)& F_{n}(-N,N) &G_{n}(N-3,N-1)\\
(1-\epsilon Q_n Q_{-n}^*)G_{-n}^{*}(N-1,N-1)&G_{1-n}^{*}(-N,N) &-\epsilon F_{1-n}^{*}(N-3,N-1)\end{vmatrix}}
{\begin{vmatrix}F_{n+1}(N-2,N) & G_{n}(N-2,N)\\
(1-\epsilon Q_n Q_{-n}^*)G_{-n}^{*}(N-2,N) & -\epsilon F_{1-n}^{*}(N-2,N)\end{vmatrix}}, \label{A6}  \\
R_{n}^{[N]}= &\frac{\epsilon Q_{-n}^*\tau_n(N-1,N-1;N,N-2)+(-1)^N\tau_n(N-1,N-3;N,N)}
{(-1)^N\tau_n(N,N-2;N-1,N-1)}\notag\\
= &\frac{\begin{vmatrix}F_{n+1}(N-2,N-2)&  G_{n+1}(1-N,N-1) &G_{n}(N,N-2) \\
(1-\epsilon Q_n Q_{-n}^*)G_{-n}^*(N-2,N-2)& -\epsilon F_{-n}^*(1-N,N-1)  &-\epsilon F_{1-n}^{*}(N,N-2)\end{vmatrix}}{(-1)^N
\begin{vmatrix}F_{n+1}(N-1,N-1)& G_{n}(N-1,N-1) \\
(1-\epsilon Q_nQ_{-n}^*)G_{-n}^*(N-1,N-1)& -\epsilon F_{1-n}^*(N-1,N-1)\end{vmatrix}}.  \label{A7}
\end{align}}}
Taking complex conjugate  and changing $n\ra -n$ for $Q_n^{[N]}$ in Eq.~\eref{A6}, one can check that $R^{[N]}_n = \epsilon Q^{*[N]}_{-n} $ is exactly satisfied.

\begin{thebibliography}{99}
\bibitem{bender1}
C.\ M.\ Bender and S.\ Boettcher, {Phys.\ Rev.\ Lett.} {\bf 80},
5243 (1998).

\bibitem{NO}
Z.\ H.\ Musslimani, K.\ G.\ Makris, R.\ El-Ganainy and D.\ N.\
Christodoulides, Phys. Rev. Lett. {\bf 100}, 030402 (2008).

\bibitem{crystal}
S. Longhi, 
{Phys. Rev. Lett.}  {\bf 103}, 123601 (2009).      

\bibitem{Markum}
H. Markum, R. Pullirsch and T. Wettig,
{Phys. Rev. Lett.} {\bf 83}, 484 (1999). 

\bibitem{BEC}
H. Cartarius and G. Wunner, {Phys. Rev. A} {\bf 86}, 013612 (2012).



\bibitem{Ablowitz1}
M.\ J.\ Ablowitz and Z.\ H.\ Musslimani, {Phys.\ Rev.\ Lett.} {\bf110}, 064105 (2013).


\bibitem{Sarma}
A.\ K.\ Sarma, M.\ A.\ Miri, Z.\ H.\ Musslimani and D.\
N.\ Christodoulides, {Phys.\ Rev.\ E}\ {\bf89}, 052918 (2014).


\bibitem{Khare}
A.\ Khare and A.\ Saxena, {J.\ Math.\ Phys.}\ {\bf 56}, 032104 (2015).



\bibitem{LiXuPRE}
M.\ Li and T.\ Xu, {Phys. Rev. E}\ {\bf91}, 033202 (2015).

\bibitem{XuLiJPSJ}
T.\ Xu and M.\ Li, arXiv:1503.02254.


\bibitem{Sarma2}
S. K. Gupta and A. K. Sarma, 
{Commun.\ Nonlinear Sci.\ Numer. Simulat.} {\bf 36}, 141 (2016). 


\bibitem{Yan1}
Z.\ Yan, 
{Appl.\ Math.\ Lett.} {\bf47}, 61 (2015). 



\bibitem{Fokas}
A. S. Fokas, {Nonlinearity} {\bf29}, 319 (2016). 


\bibitem{Ablowitz3}
M. J. Ablowitz and Z. H. Musslimani, {Phys.\ Rev.\ E}\ {\bf90}, 032912 (2014).

\bibitem{SW}
D.\ S.\ Wang and X.\ Q. Wei, Appl. Math. Lett. {\bf 51}, 60 (2016);
L. Wang, C. Geng, L. L. Zhang and Y. C. Zhao, {EPL} {\bf 108}, 50009 (2014).


\bibitem{Roguewaves}
R.\ Guo, Y. F. Liu, H. Q. Hao and F. H. Qi, {Nonlinear Dyn.} {\bf 80}, 1221 (2015);  
L. C. Zhao and J. Liu, {Phys. Rev. E}  {\bf 87}, 013201 (2013);  L. M. Ling, B. L. Guo and L. C. Zhao, {Phys. Rev. E} {\bf 89}, 041201 (2014).

\bibitem{Zhu}
L.Y. Ma, Z.N. Zhu, {Appl. Math. Lett.} {\bf 59} (2016) 115.

\bibitem{DAL}
Y.\ Ohta and J. K. Yang, J. Phys. A {\bf 47}, 25520 (2014).


\end{thebibliography}
\end{document}